# Time-resolving magnetic scattering on rare-earth ferrimagnets with a bright soft-X-ray high-harmonic source

G. Fan[1,2], K. Légaré[2], V. Cardin[2], X. Xie[1,9], E. Kaksis[1], G. Andriukaitis[1], A. Pugžlys[1], B. E. Schmidt[3], J.P. Wolf[4], M. Hehn[7], G. Malinowski[7], B. Vodungbo[5], E. Jal[5], J. Lüning[5], N. Jaouen[6], Z. Tao[8*], A. Baltuška[1], F. Légaré[2] and T. Balčiūnas[1,4]

[1] *Institute of Photonics, TU Wien, Gusshausstrasse 27/387, Vienna, Austria*
[2] *Institut National de la Recherche Scientifique, Varennes, Quebec J3X1S2, Canada*
[3] *few-cycle, Inc., 2890 Rue de Beaurivage, Montreal, Quebec H1L 5W5, Canada*
[4] *GAP-Biophotonics, Université de Genève, 1205 Geneva, Switzerland*
[5] *Sorbonne Université, CNRS, Laboratoire de Chimie Physique - Matière et Rayonnement, LCPMR, 75005 Paris, France*
[6] *Synchrotron SOLEIL, L'Orme des Merisiers, 91192 Gif-sur-Yvette, France*
[7] *Institut Jean Lamour, CNRS UMR 7198, Université de Lorraine, 54000 Nancy, France.*
[8] *State Key Laboratory of Surface Physics, Department of Physics, Fudan University, Shanghai 200433, People's Republic of China*
[9] *SwissFEL, Paul Scherrer Institute, 5232 Villigen PSI, Switzerland*

* Email: Zhenshengtao@fudan.edu.cn

## Abstract

**We demonstrate the first time-resolved X-ray resonant magnetic scattering (tr-XRMS) experiment at the *N* edge of Tb at 155 eV performed using a tabletop high-brightness high-harmonic generation (HHG) source. In contrast to static X-ray imaging applications, such optical-pump—X-ray-probe studies pose a different set of challenges for the ultrafast driver laser because a high photon flux of X-rays resonant with the *N* edge must be attained at a low repetition rate to avoid thermal damage of the sample. This laboratory-scale X-ray magnetic diffractometer is enabled by directly driving HHG in helium with terawatt-level 1 μm laser fields, which are obtained through pulse compression after a high-energy kHz-repetition-rate Yb:CaF$_2$ amplifier. The high peak power of the driving fields allows us to reach the fully phase-matching conditions in helium, which yields the highest photon flux (>2x10$^9$ photons/s/1% bandwidth) in the**

**100-220 eV spectral range, to the best of our knowledge. Our proof-of-concept tr-XRMS measurements clearly resolve the spatio-temporal evolution of magnetic domains in Co/Tb ferrimagnetic alloys with femtosecond and nanometer resolution. In addition to the ultrafast demagnetization, we observe magnetic domain expansion with a domain wall velocity similar to that induced by spin transfer torque. The demonstrated method opens up new opportunities for time-space-resolved magnetic scattering with elemental specificity on various magnetic, orbital and electronic orderings in condensed matter systems.**

Spontaneous emergence of magnetic orders in nanoscale and mesoscale structures has been widely observed and plays an important role in a variety of macroscopic phenomena in magnetic materials. Such heterogeneity of magnetic states in an otherwise spatially homogeneous material is a result of a complex interplay between electron spins and other degrees of freedom (electron orbitals and lattice). Driven by optical excitation, spin angular momentum can be transferred between neighboring magnetic nanoregions with different constituent magnetic elements, facilitating magnetization reversal on sub-picosecond timescales in rare-earth–transition-metal ferrimagnetic alloys and multilayers[1–3]. Such ultrafast spin transfer promises potential applications in future data storage and spintronic devices functioning on the picosecond timescale, but its underlying mechanisms are still under debate[4–6]. To further advance the field, it is essential to develop microscopic methods that can map optically-induced magnetization reordering processes with elemental specificity, and at their characteristic sub-picosecond temporal and nanometer spatial scales.

High-brightness HHG[7–9] light sources enable tr-XRMS as an excellent laboratory-scale tool for studying nanoscale magnetic dynamics. Owing to its high spatial coherence, HHG radiation has been widely used to provide high-quality images of nanostructures with a large-scale view[10–12]. In addition to the outstanding spatial properties, HHG pulses enable femto-to-attosecond temporal resolution[13,14]. The broad spectral bandwidth of HHG spans the characteristic *M* and *N* absorption edges of the transition-metal (TM) and rare-earth (RE) elements that exhibit magneto-optical activity. X-ray magnetic circular dichroism (XMCD) at these edges[15] provides access to an element-specific[16,17] mapping of magnetic states and allows following their evolution in time. Compact tabletop HHG sources addressed in this Letter favorably differ in size and complexity from facility-scale sources such as free-electron lasers (FEL) [3,18] and synchrotrons with femtosecond slicing technology[19].

Compared to other spectroscopic methods, resonant magnetic scattering requires orders of magnitude higher photon flux because of the extremely low, ~$10^{-5}$ scattering cross-section of magnetic structures in the XUV spectral range[20], which imposes a major challenge for the applications of HHG radiation in scattering experiments. To date, tr-XRMS based on tabletop HHG sources has only been carried out on 3*d* TM ferromagnets (iron, cobalt and nickel) and a few multilayer structures[20–22], by covering the easily accessible *M* absorption edges of the TM elements at ~60 eV. To image the laser-induced nanoscale spin transfer in ferrimagnets, such as TbFeCo[23] and GdFeCo[1,3], it is thus crucial to further extend HHG photon energy and cover the *N* edges of 4*f* RE ferromagnets (Gd (~148 eV), Tb (~155 eV) and Dy (~153 eV)). One the one hand, it is possible to extend the phase-matched HHG cutoff with longwave mid-IR driver pulses from optical parametric amplifiers (OPA)[24–28]. On the other hand, scaling the wavelength of the driver pulse reduces the HHG efficiency[29] as $\lambda_L^{-5.5 \pm 0.5}$. Moreover, the low efficiency of optical frequency conversion in an OPA (10-20%) further limits the HHG flux. In contrast, high-brightness HHG can be achieved using 1030 nm pulses from high-repetition-rate Yb fiber lasers[30] by applying a tight-focusing geometry and a high-ionization-potential target gas. Nevertheless, because the phase-matching pressure of the target gas scales inversely with the square of the beam diameter[31], it is very challenging to fulfill phase-matching conditions, precluding applications of such sources in the >150 eV region.

In this work, we demonstrate the first tr-XRMS on $Co_{0.88}Tb_{0.12}$ ferrimagnetic alloys with high-brightness HHG radiation covering the *N* edge of Tb at 155 eV. This ultrafast magnetic diffractometer is enabled by HHG up to 220 eV with the highest brightness (>2×10$^9$ photons/s/1% bandwidth (BW)) ever reported in the literature. This is accomplished by directly driving the HHG process in helium with 1030-nm 0.3-TW-peak-power femtosecond laser pulses resulting from hollow-core fiber (HCF) pulse compression of a high-energy Yb:CaF$_2$ laser system. The combination of optimum driver wavelength and high peak power boosts the HHG flux at 155 eV for tr-XRMS. In this work, we prove this assertion both experimentally and theoretically by comparing various generation schemes employing different wavelengths, gases and pressures. Our 220 eV, high-brightness HHG source with a-few-kHz repetition rate is especially suitable for time-resolved soft X-ray spectroscopy and imaging experiments on solids, for which high-repetition-rate sources could cause irreversible thermal damage induced by the optical pump. From the tr-XRMS measurements, we observed the laser-induced evolution of magnetic domains in a $Co_{0.88}Tb_{0.12}$ ferrimagnetic alloys on the femtosecond time scale with domain changes on the nanometer scale, revealing different dynamics compared to previous work[18,21].

The experimental setup is illustrated in Fig. 1. The fundamental ~220 fs, 1030 nm pulses from a Yb:CaF$_2$ amplifier were first compressed by a post-pulse-compressor, consisting of a HCF and a set of chirped mirrors. The implemented HCF is 3-m-long and has a large core diameter (1mm), enabling a compression ratio of ~10 for the high-energy (~11 mJ) pulses. The post-compressed pulse has a peak power of 0.3 TW (~25 fs, 8 mJ), which is the key to achieving efficient laser-like HHG (Fig. 2c) up to 220 eV in a 20 mm helium-filled gas cell with adjustable backing pressure. The harmonic spectrum is characterized with a soft X-ray spectrometer (see Methods). As shown in Fig. 2a and e, the change of the pulse duration from 220 fs to 25 fs extends the HHG cutoff from 150 eV to 220 eV. This result can be explained by the suppression of ionization for a shorter driver pulse[32].

To generate high-brightness HHG radiation, it is essential to reach the optimal phase matching conditions. In Fig. 2b, we plot the total spectrum intensity for photon energy >100 eV as a function of the backing pressure of helium, at a laser peak intensity of $6.5\times10^{14}$ W cm$^{-2}$. The HHG intensity grows quadratically at low backing pressure, followed by a saturation of the signal at a backing pressure of 200 mbar. By further increasing the pressure, the spectral intensity decreases due to the absorption of the generated harmonics within and after the generation volume. This observation is reproduced with high fidelity in our *ab-initio* simulation based on the strong-field approximation (see Fig. 2c and Methods) and demonstrates that the HHG generated in helium using our approach is fully phase-matched and absorption-limited in intensity (see Supplementary). Compared to previous experiments[30,31], the high peak intensity of the driving laser here allows us to use a loose focusing geometry and relax the requirements on the phase-matching pressure.

We also experimentally compare the conversion efficiency of our approach (*i*) 1 μm in helium with two other generation schemes based on the use of OPAs: (*ii*) 1.5 μm in neon and (*iii*) 2.4 μm in argon. Since our target is to optimize the HHG brightness at 220 eV, we adjust the peak power, pulse duration and focusing geometry of the driving laser in every case to reach the same cutoff energy. For gases with a lower ionization potential, it is essential to use a longwave driver pulse to stay below the critical ionization level[33]. The backing pressure of the gas medium is also optimized to obtain the phase-matched and absorption-limited HHG. The spectrometer efficiency, filter transmission and the input pulse energies at different conditions are taken into account to extract the conversion efficiency right after the gas cell (see Methods). The experimentally measured conversion efficiencies for the three different approaches are plotted in Fig. 3a, showing that our method ((*i*) 1μm in helium) yields the highest conversion efficiency throughout the 100~200 eV range of interest, which is supported by our simulations shown in Fig. 3b.

For the phase-matched HHG, the absorption-limited conversion efficiency can be described as $\xi_q = \lambda_L^{-n} \left|\frac{A_q}{\sigma}\right|^2$ [34], where $A_q$ is the amplitude of the single-atom recombination cross-section at the harmonic frequency $\omega_q$, $\lambda_L^{-n}$ represents the wavelength scaling due to the electron wave packet diffusion during its free-space excursion, with $n = 5.5 \pm 0.5$[29], and $\sigma$ is the X-ray absorption cross-section. As shown in Fig. 3c, the longest electron wave packet excursion occurs for argon with $\lambda_L$=2.4 µm, which leads to significant reduction of the recombination probability for the case (*iii*)[9]. In contrast, the single-atom responses for (*i*), 1 µm in helium, and (*ii*), 1.5 µm in neon, are similar since the stronger wave packet diffusion for the longer $\lambda_L$ is compensated by the larger recombination cross-section associated with the larger ionic core of neon in case (*ii*). Summarizing the above discussion of microscopic single-atom response, the expected HHG efficiency is similar for both helium and neon. Nevertheless, macroscopic propagation changes the situation in favor of helium because its X-ray absorption cross-section ($\sigma$) is one order of magnitude lower than neon at the energy of 200 eV (Fig. 3a and b).

For a fixed driver pulse duration, this simple model suggests a straightforward recipe for reaching the highest flux at a target X-ray photon energy located in a resonance-free plateau region of the harmonic spectrum in the vicinity of the cutoff. The highest efficiency is achieved using helium driven by the shortest laser wavelength capable of reaching the corresponding semi-classical cutoff, given that the laser pulse intensity is sufficient to sustain a fully phase-matched HHG regime.

In Fig. 3d, we show that, owing to the high conversion efficiency for 1 µm driving HHG in helium, we achieve the highest flux of $2\times10^9$ photons/s/1%BW at 200 eV. Unlike most of the previous experiments summarized in Fig. 3d, in which $\lambda_L$-dependent cutoff extension was studied, our method is free of additional energy loss due to the absence of parametric frequency conversion. Reviewing HHG results with direct laser driving at different wavelengths, it must be noted that with 800 nm laser pulses from a Ti:Sapphire amplifier, it was possible to extend the cutoff beyond 200 eV by employing sub-10-fs driver pulses[35] or using quasi-phase matching techniques[36]. However, in these situations, ionization-induced-phase-mismatch quickly outruns the dispersion contribution of the neutral atoms, making the macroscopic phase matching very challenging[37]. By contrast, in our experiments with 1-μm driver pulses, we significantly suppressed the ionization of helium thus facilitating the phase matching. The estimated ionization fraction was below 0.38%, while the critical ionization of helium is 0.4%.

The high-brightness HHG up to 220 eV generated using our new method allows us, for the first time, to carry out the tr-XRMS measurements on the ferrimagnetic alloy $Co_{0.88}Tb_{0.12}$. The sample was a 50 nm film grown on a $Si_3N_4$ membrane, exhibiting an out-of-plane magnetic anisotropy with a stripe domain structure (Fig. 1). A concave multilayer mirror focuses a portion of the harmonic beam on the ferrimagnetic sample and selects a 5-eV-wide spectral bandwidth covering the *N* edge of Tb at 155 eV (see Fig. 2d). Due to the XMCD effects, the alternating oppositely-magnetized domains serve as a diffraction grating for the incident linearly polarized soft X-ray beam, giving rise to the ±1$^{st}$-order diffraction peaks in the far field, as illustrated in Fig. 4a. The magnetic domain structure resolved using magnetic-force microscopy (MFM) of the same sample is plotted in Fig. 4b. Correspondingly, the Fourier transform (FT) of the real-space stripe-like structure yields a diffraction pattern consisting of two well-defined diffraction spots, the momentum transfer (*k*) of which is consistent with that obtained through XRMS using the bright harmonic beam, as shown in Fig. 4c.

The ultrafast dynamics in the magnetic material is induced by 1.5 µm, 80 fs laser pulses obtained from an OPA driven by the same Yb driver laser, and is probed by soft X-ray pulses arriving at the sample with a time delay $t_d$. The repetition rate of the laser was intentionally reduced from 2 kHz to 500 Hz in order to prevent thermal damage of the sample by the accumulated heat from the pump excitation. This demonstrates the need for high pulse energy laser systems operating at low repetition rate for driving tabletop X-ray sources for applications in solid-state physics, e.g. thin films with low heat dissipation and materials with slow recovery.

The domain magnetization amplitude $M$ can be measured by the square root of the diffraction intensity, while the spatial evolution of the magnetic domains is revealed by the change of the momentum transfer ($\Delta k$)[21]. As shown in Fig. 4d, with a pump fluence of 8 mJ/cm$^2$, the intensity of the diffraction peaks is suppressed by ~70%, which corresponds to demagnetization up to ~50%. The demagnetization of the sample exhibits two timescales: an ultrafast demagnetization process quickly suppresses ~10% of the sample magnetization in the first 680 fs, followed by a slow demagnetization process in ~18 ps. The two-step process is consistent with the "Type II" demagnetization dynamics previously observed in pure Tb and $Gd_{1-x}Tb_x$ alloys[6,38]. More interestingly, we find that the momentum transfer ($k$) is reduced by ~3% in ~10 ps after pump laser excitation, indicating an expansion of the periodicity of the magnetic domains perpendicular to the stripe-structures (Fig. 4e, x direction). As shown in Fig. 4e, the decay of $\Delta k/k$ is much slower than the sample demagnetization ($M(t)/M_0$) and can be approximated as a linear decrease as a function of time. The slope of the change yields a velocity for the domain expansion of ~750 m/s. Very interestingly, this velocity coincides with the domain-wall velocity under the current-induced spin-transfer torque, with a current density $>3\times10^{12}$ A/m$^2$ [39]. Indeed, it has been shown that such a high current density can possibly be created with similar pump fluences[40] as in our experiment, leading to rearrangement of the domain pattern on picosecond timescales. We note that the domain wall velocity here is much smaller compared to the velocity observed in the CoPt multilayers[18], indicating a very different driving mechanism in 4*f* RE ferrimagnets materials compared to 3*d* metals.

In conclusion, we demonstrated the first time-resolved X-ray magnetic diffractometer based on tabletop high-brightness HHG source reaching the *N* edge of Tb at 155 eV. The measured time-dependent diffraction patterns allow us to extract the ultrafast demagnetization as well as the temporal evolution of magnetic domains with nanometer spatial resolution. Seemingly counter-intuitively, the contrast of pump-probe measurements was improved by decreasing the repetition rate, thus allowing us to increase the pump fluence to quench efficiently the magnetization while keeping the average pump power low enough to prevent thermal damage. This indicates the urgent necessity to develop high-energy sub-kHz-repetition-rate laser sources delivering ultrashort pulse duration. These future technologies will enable high flux harmonic sources, despite lower repetition rate, to perform high-resolution spatio-temporal imaging of various magnetic, orbital and charge orderings of condensed matter.

**Figures 1-4**

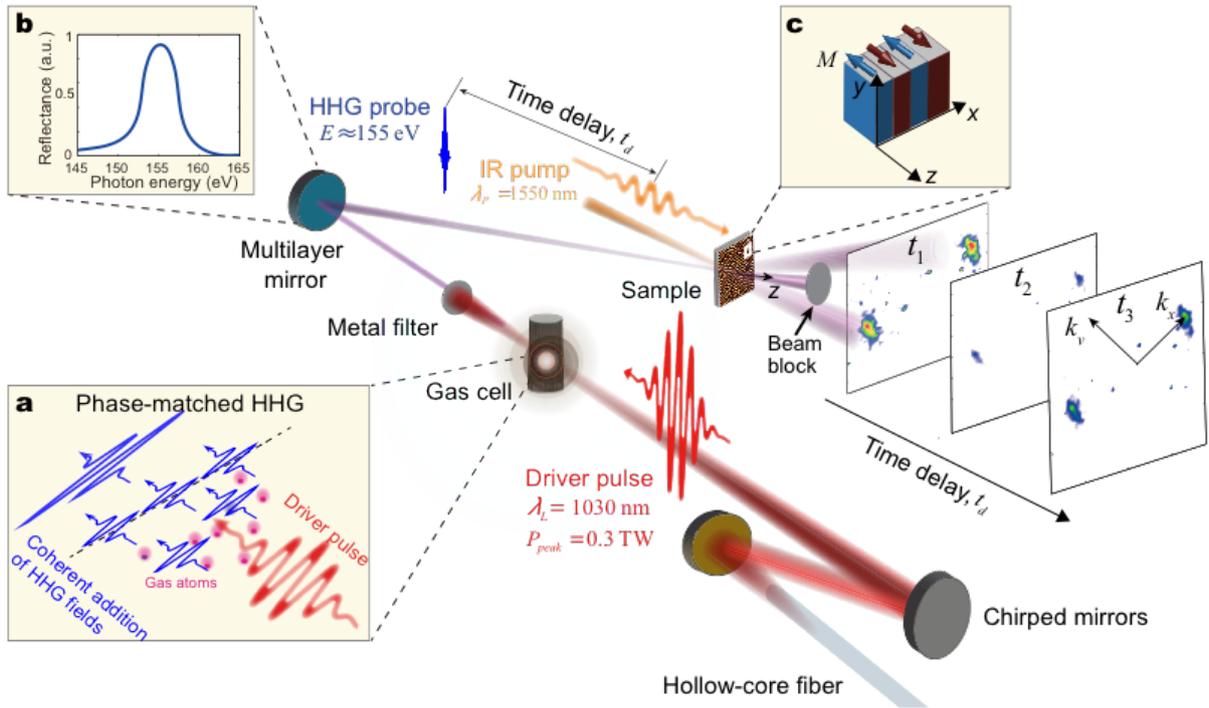

**Fig. 1 The schematics of the Tr-XRMS experiment setup**. High-brightness high-order harmonics are generated in a gas cell filled with helium and driven by the compressed pulses in the full phase matching conditions (inset a). The harmonic beam is then focused by a multilayer mirror, which selects the harmonic spectrum around 155 eV, corresponding to the *N* edge of Tb (inset b). Tr-XRMS experiments are carried out in a pump-probe geometry, by exciting the sample with 1550 nm IR pulses before HHG probe pulses arrive. The sample is a 50 nm $Co_{0.88}Tb_{0.12}$ thin film deposited on a $Si_3N_4$ membrane with an out-of-plane magnetic anisotropy (inset c). The spatio-temporal evolution of the magnetic domains is measured with the time-resolved diffraction patterns from the harmonic beam, recorded with a charge-coupled-device (CCD) camera as a function of the delay time $t_d$.

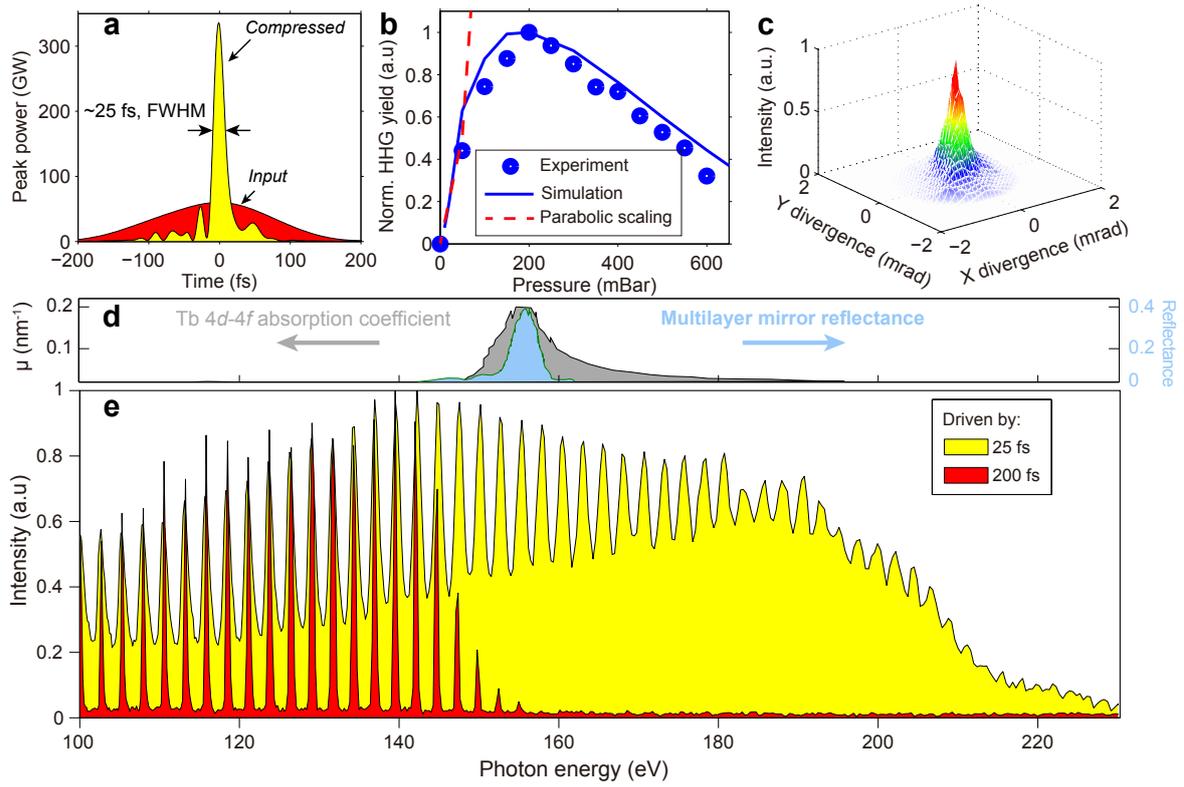

**Fig. 2 Driver pulse compression and HHG characterization.** (a) Temporal characterization of uncompressed and compressed driver pulses with the second-harmonic-generation frequency-resolved optical gating (SHG-FROG). (b) Measured and simulated HHG signals as a function of the helium backing pressure with a peak power of ~0.4 TW at the wavelength of 1030 nm. (c) The beam profile of the narrow band harmonic beam after the multilayer mirror with the central photon energy of 155 eV. (d) Absorption cross section of Tb near its *N* edge and the reflectance of the X-ray multilayer mirror. (e) High-order harmonic spectra driven by the uncompressed (200 fs) and compressed (~25 fs) pulses, respectively

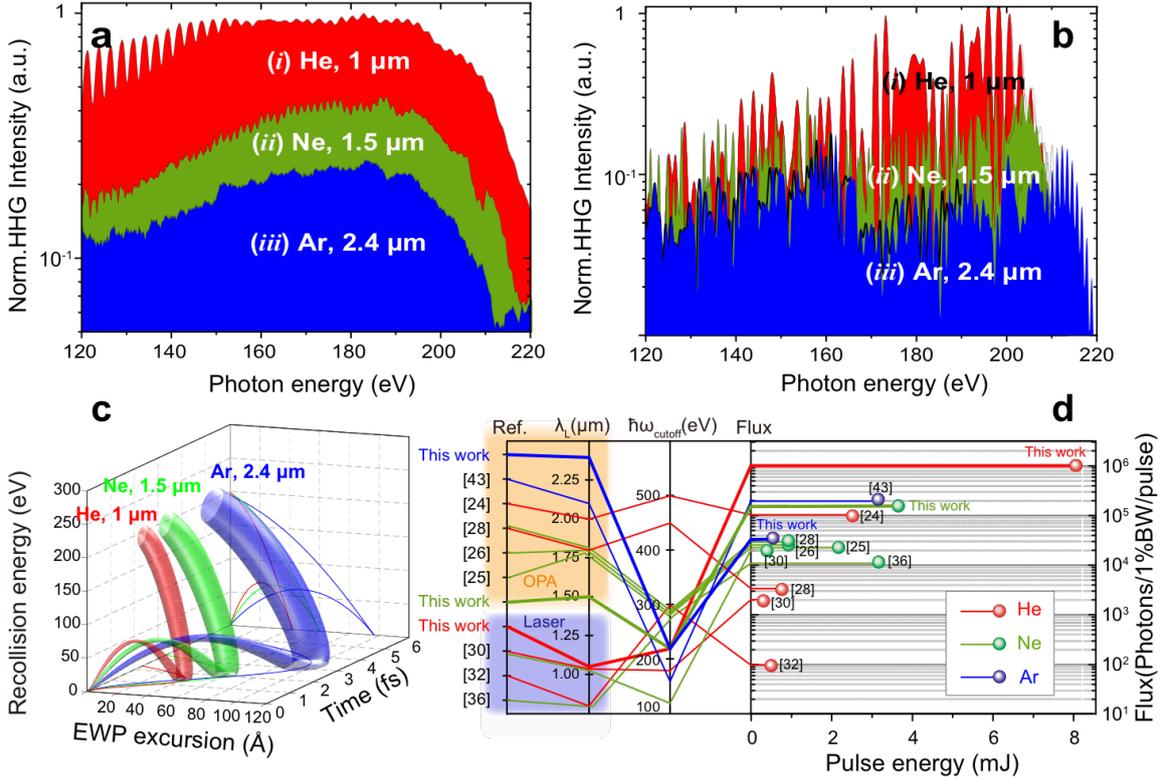

**Fig 3. High brightness of HHG directly driven by compressed 1μm laser pulses.** (a) Measured and (b) simulated conversion efficiencies of HHG in the 100~200 eV spectral range generated in (*i*) helium with 1 μm, in (*ii*) neon with 1.5 μm and in (*iii*) argon with 2.4 μm driving wavelengths. The peak power, pulse duration and focusing geometry are adjusted to yield the same cut-off energy of the spectra. The conversion efficiencies in (a) and (b) are all normalized to the conversion efficiency in case (*i*) at ~200 eV. c) The simulated cutoff trajectories for the three cases shown in (a) and (b). The radii of the trajectories linearly increase with the excursion time, symbolizing wave packet spreading. The more efficient single-atom response for 1 μm in helium is due to reduced electron wave packet (EWP) spread caused by the shorter wavelength driving field. (d) Overview of the experimentally generated HHG flux in photons per shot per 1%BW above 100 eV in helium (red)[24,28,30,32], neon (green)[25,26,28,30,36] and argon (blue)[43]. The driving field wavelength ($\lambda_L$) is illustrated by marker colors.

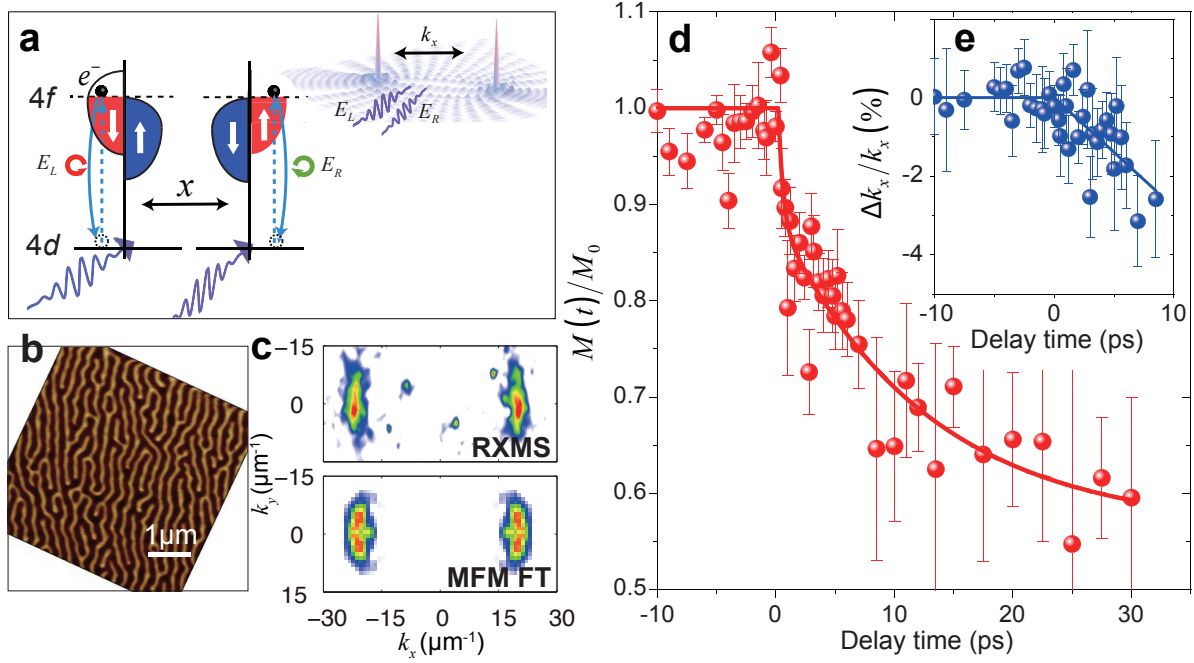

**Fig. 4 tr-XRMS measurements of CoTb sample at the *N* edge of Tb.** (a) Illustration of the physical mechanism of XRMS using linearly polarized XUV light. (b) Magnetic domain structure of the same CoTb sample measured using a magnetic force microscope (MFM). (c) The diffraction pattern corresponding to the magnetic domain. The upper panel plots the experimentally measured diffraction pattern using XRMS as a function of momentum transfers $k_x$ and $k_y$. The lower panel shows the Fourier transform (FT) of MFM image in (b), which yields the diffraction peaks with the momentum transfer consistent with the XRMS measurement. (d) The variation of averaged magnetization ($M(t)$) as a function of the pump-probe delay time, extracted from the time-dependent variation of the XRMS diffraction patterns. The magnetization is normalized to the ground-state magnetization $M_0$. (e) The change of momentum transfer in the x direction in percentage ($\Delta k_x/k_x$) as a function of the delay time.

## Methods

**Setup for XRMS measurement**

**HHG generation and characterization**

High-brightness HHG beam by 1030 nm femtosecond pulses with TW-level peak power, which was directly obtained through pulse compression of ~200 fs, 11 mJ, 1030 nm pulses from a Yb:CaF$_2$ amplifier. The pulse compression was achieved in a 3-meter long Ar-filled HCF and the spectrally broadened pulses were compressed by a set of chirped mirrors down to ~25 fs pulse duration with an efficiency of >75%. To achieve a high compression factor (~10), we employed a stretched waveguide technique[44]. The specially designed flexible HCF has a 1mm inner diameter with ~300-µm-thick fused silica cladding surrounded by a polymer layer.

In order to find the optimal conditions for generating HHG in the 100-220 eV spectral region, we compare the HHG conversion efficiency of HHG generated in helium with 1μm driving fields with generation schemes using OPAs (1.5 μm in neon and 2.4 μm in argon). The 2.4 μm light was generated by a 3-stage OPA with a pulse energy of ~0.8 mJ, while 1.5 μm light was generated by a 4-stage OPA with a pulse energy of 3.7 mJ. The phase matching conditions in argon and neon were optimized correspondingly by adjusting the focusing geometry as well as the gas-cell length to achieve fully phase-matching conditions, respectively.

HHG spectrum was measured with an XUV spectrometer, which is based on a flat-field grating with a nominal groove number of 1200 lines/mm and an X-ray CCD camera (Andor Newton 920). The image of the 50-μm-wide slit is imaged direct onto the camera through the concave grating. The absolute photon flux of HHG generated in helium with 1μm fields was estimated, on the other hand, in the diffraction experimental geometry by considering the quantum efficiency of the CCD camera, the transmittance of the filter and the reflectance of the multilayer mirror in the diffraction experimental geometry. The fluxes of HHG generated by the other two generation schemes were then calculated relatively (see Supplementary).

**TDSE Simulation**

In order to obtain simulated harmonic spectra from the interaction of noble gases with a strong laser field, we numerically solved three-dimensional time-dependent Schrödinger equation in velocity gauge with single active electron approximation [45],

$$i\frac{\partial}{\partial t}\Psi(\mathbf{r},t) = \left[\frac{(\mathbf{p}-\mathbf{A}(t))^2}{2} - V(\mathbf{r})\right]\Psi(\mathbf{r},t),$$

where $V(\mathbf{r})$ is the potential and $\mathbf{A}(t) = \sqrt{I_0}/\omega_0 \, e^{-1.38\frac{t^2}{\tau^2}} \cos(\omega_0 t)$ is the vector potential of the laser field with a Gaussian envelope, a peak intensity of $I_0$, a center frequency of $\omega_0$ and a pulse duration of $\tau$. In these calculations, we employed the pseudospectral method with Tong-Lin models[45] of the helium, neon and argon atom. Afterwards, single-atom harmonic spectra were calculated from the Fourier transform of the dipole acceleration with the simulated electron wave functions. The laser wavelengths and the pulse durations for the simulations were the same as those in the corresponding measurements, while the pulse intensities were chosen to generate the corresponding cutoff energies.

**Sample magnetization**

Probing of magnetization state via small-angle X-ray magnetic scattering using linearly polarized X-ray pulses rely on preparation of periodically magnetized magnetic domain structure of the sample. The aligned magnetic stripe domains can be achieved using a magnetization procedure described in [46]. A thin-layer ferromagnetic sample with saturated out-of-plane magnetic moment forms a "labyrinth" domain state with a typical domain size that is dependent on the thickness of the sample. In order to achieve the aligned "stripe" domain state, the sample is de-magnetized using a strong magnetic field along an in-plane axis.

Acknowledgements:

Z. Tao gratefully acknowledge support from the National Natural Science Foundation of China (grant no. 11874121) and the Youth Thousand Talent program of China. T.B. acknowledges funding from the EU H2020 resarch and innovation programme under the Marie Sklodowska-Curie grant agreement No 798176. X. Xie acknowledges funding from Austrian Science Fund (FWF) P30465-N27.